\documentclass[11pt, journal,draft,onecolumn]{IEEEtran}
\usepackage{amsmath}
\usepackage{txfonts}
\usepackage{setspace}
\usepackage{enumerate}

\usepackage{stmaryrd}
\usepackage{amssymb}
\usepackage{mathrsfs}
\usepackage{amsfonts}  
\usepackage{flushend}
\usepackage[dvips,final]{graphicx}
\usepackage[dvips]{color}
\usepackage{epsfig,amssymb}
\newcommand{\beq}{\begin{equation}}
\newcommand{\eeq}{\end{equation}}
\newcommand{\beqn}{\begin{eqnarray}}
\newcommand{\eeqn}{\end{eqnarray}}

\begin{document}


\newpage

\begin{center}

{\Large\bf CSSF MIMO RADAR: Low-Complexity   Compressive Sensing
Based MIMO Radar That Uses Step Frequency   \footnote{ This work was
supported by the Office of Naval Research under Grant
ONR-N-00014-07-1-0500, and the National Science Foundation under
Grants  CNS-09-05398 and CNS-04-35052. }}

\bigskip
{\it Yao Yu and Athina P. Petropulu} \\
          Department of Electrical \& Computer Engineering,
          Drexel University, Philadelphia, PA 19104\\
          \medskip
          {\it H. Vincent Poor}\\
          School of Engineering and Applied Science,
          Princeton University, Princeton, NJ 08544
\end{center}

\bigskip

\begin{abstract}

A new approach is proposed, namely  CSSF MIMO radar, which  applies
the technique of step frequency (SF) to compressive sensing (CS)
based multi-input multi-output (MIMO) radar. The proposed approach
enables high resolution range, angle and Doppler estimation, while
transmitting narrowband pulses. The problem of  joint
angle-Doppler-range estimation is first formulated to fit the CS
framework, i.e., as an $\ell_1$ optimization problem. Direct
solution of this problem  entails high complexity as it employs a
basis matrix whose construction requires discretization of the
 angle-Doppler-range space. Since high resolution requires fine space discretization, the complexity of
 joint range, angle and Doppler estimation can be prohibitively high.
 For the case of slowly moving targets, a technique is proposed that achieves significant complexity reduction
 by successively estimating angle-range and
Doppler in a decoupled fashion and by employing initial estimates
obtained via matched filtering    to further reduce the space that
needs to be digitized. Numerical results show that the combination
of CS and SF results in a MIMO radar system that has superior
resolution and requires far less data as compared to a system that
uses a matched filter  with SF.


\end{abstract}
\section{Introduction}

Benefiting from  the idea of multiple-input multiple-output (MIMO)
communication systems, MIMO radar systems have received considerable
attention in recent years. A MIMO radar
\cite{Fishler:04}-\cite{Li:07} although having some similarities
with phased-array radar,  is different in  that it transmits
multiple
 waveforms from its antennas,  and in general  offers a more
flexible configuration. A MIMO radar
 with widely separated antennas \cite{Haimovich: 08} views a
target from different decorrelated directions and thus enjoys
 spatial
diversity. For  MIMO radar with colocated antennas \cite{Stoica:
07m}\cite{Chen:081},  independent waveforms enable superior angular
resolution as compared to phased-array radar.  We focus on the
colocated type of MIMO radar throughout this paper.

Compressive sensing (CS) is a relatively recent development
\cite{Donoho:06}-\cite{Romberg:08}  and has already been applied
successfully in diverse fields such as image processing
 and wireless communications.   The theory of CS states that a
$K$-sparse signal vector $\mathbf{x}$ of length $N$ can be recovered
exactly with high probability based on $\mathcal{O}(K\log N)$
measurements via $\ell_1$-optimization. Let $\bf \Psi$  denote the
 \emph{basis matrix} that spans the space in which the signal is
sparse,
 and  $\bf\Phi$ be the {\it measurement matrix} that is used to linearly compress
the signal. The $\ell_1$-optimization problem  is formulated as
follows:
\begin{eqnarray}
\min\|\mathbf{s}\|_1,\ \  s.t. \ \text{to}\ {\bf
y}=\mathbf{\Phi}{\bf x} =\mathbf{\Phi \Psi} {\bf s} \label{CS}
\end{eqnarray}
where $\mathbf{s}$ is an $N\times 1$ sparse vector with $K$
principal elements while the remaining elements can be ignored;
$\mathbf{\Phi}$ is an $M\times N$ matrix with $M\ll N$.   The
product $\mathbf{\Phi\Psi}$ is referred to as the {\it sensing matrix}.
The uniform uncertainty principle (UUP)
\cite{Candes:08}\cite{Candes:062}
 indicates that if every set of  sensing
matrix columns with cardinality less than the sparsity of the signal
of interest    is approximately orthogonal, then the sparse signal
can be exactly recovered with high probability. This  implies that
$\mathbf{\Phi}$ is incoherent with $\mathbf{\Psi}$. Throughput this
paper, we will refer to the solution of a problem of the form
({\ref{CS}}) as the {\it CS approach}.

By exploiting the sparsity of radar signals in various spaces,   CS
has been applied to radar systems
\cite{Baraniuk:07}-\cite{Herman:08} and to MIMO radar
\cite{Petropulu:08}-\cite{Yu:09_tsp}. In \cite{Chen:08} and
\cite{Strohmer:Asilomar09}, a uniform linear array was considered as
a transmit and receive antenna configuration and the CS approach was
applied using a submatrix of the identity matrix as the measurement
matrix.
 Bounds on the achievable range and
azimuth resolution and the number of recoverable targets were
derived in \cite{Strohmer:Asilomar09}.
 In \cite{Petropulu:08} and \cite{Yu:09_tsp}, the authors
proposed a CS-based MIMO radar system implemented on a small scale
network. According to \cite{Yu:09_tsp}, randomly located network
nodes, each equipped with a single antenna, serve as transmit and
receive antenna elements. The transmit nodes transmit periodic
pulses. The receive nodes forward their compressively obtained
measurements to a fusion center. Exploiting the sparsity of targets
in the
 angle-Doppler space, an $\ell_1$-optimization problem is formulated and
solved at the fusion center in order to extract  target angle and
Doppler information. This approach achieves the superior resolution
of MIMO radar with far fewer samples than required by conventional
approaches, which implies lower power consumption for the receive
nodes.

 The application of CS to  step-frequency
radar (SFR) \cite{Gill} was  investigated in
\cite{Gurbuz:09}-\cite{Sagar:10_ICASSP}.  SFR transmits pulse trains
of varying frequency. Thus, although the individual pulses are
relatively long in duration and are narrowband, the  transmit signal
is effectively wideband. Since  range resolution increases with the
signal bandwidth, SFR achieves high range resolution. At the same
time, SFR does not suffer from low signal-to-noise ratio (SNR) at
the receiver typically associated with wideband systems that rely on
short duration pulses.
 In
\cite{Sagar:10_ICASSP}, it was found that the CS approach
 can significantly reduce the number of pulses required by SFR to
achieve a certain resolution. A CS-based data acquisition and
imaging method was proposed in \cite{Gurbuz:09} for
stepped-frequency continuous-wave ground penetrating radars, and in
\cite{Yoon:09} CS was applied to stepped-frequency through-the-wall
radar imaging. In both cases it was  shown  that the CS approach can
provide a high-quality radar image using many fewer data samples
than conventional methods.

In this paper, we consider a more general scenario than that of
\cite{Petropulu:08} and \cite{Yu:09_tsp}. The methods of
\cite{Petropulu:08} and \cite{Yu:09_tsp} assume that the targets are
located in a small range bin and the sampling is synchronized with
the first target return. Such  assumptions do not allow for  range
estimation. In this paper, the targets can be located across several
range bins. We propose CSSF MIMO radar, an approach that applies
step frequency
 to CS-based MIMO radar.
  Two types of CSSF MIMO radar systems are considered, i.e., linear step-frequency radar (LSFR), and
random step-frequency radar (RSFR), and their effects on the CS
approach are studied. The proposed approach enables high resolution
range as well as angle and Doppler estimation.  The problem of joint
angle-Doppler-range estimation is first formulated to fit the CS
framework, i.e., as an $\ell_1$ optimization problem.
 Solving this problem entails high complexity as it employs a basis matrix whose construction requires
discretization of the
 angle-Doppler-range space. The complexity  increases with the size of the basis matrix, or equivalently, as the discretization step decreases;
 the latter step needs to be as small as possible as it sets the lower limit of resolution.
For slowly moving targets, a technique is proposed that successively
estimates angle-range and Doppler in a decoupled fashion, and
employs initial estimates obtained via a matched filter (MF) to
further reduce the space that needs to be digitized. In
\cite{Miosso: 09} and \cite{Vaswani: 09}, information on the support
of a sparse signal was used in the minimization process resulting in
complexity reduction. In our case, we do not explore the role of
initial estimates on the minimization process, as they are not
expected to be very reliable. Instead, we  use them  only as a guide
for the construction of the basis matrix. The preliminary results of
CSSF MIMO radar and the decoupled scheme were published in
\cite{Yu:Asilomar09} and \cite{Yu:ISCCSP10} which consider the same
signal models as in \cite{Petropulu:08} and \cite{Yu:09_tsp}.   This
paper extends the work of \cite{Yu:Asilomar09} and
\cite{Yu:ISCCSP10} to the general scenario aforementioned and offers
an mathematic insight into CSSF MIMO.

 The paper is organized as follows. In Section II we provide the
signal model of a CS-based MIMO radar system.  In Section III, we
introduce the proposed CSSF MIMO radar system. A decoupled scheme
for CSSF MIMO is described in Section IV. Simulation results are
given in Section V for the case of slowly moving targets. Finally,
we make some concluding remarks in Section VI.

\emph{Notation}: Lower case and capital letters in bold denote
respectively vectors and matrices.  The  expectation of a random
variable is denoted by $E\{\cdot\}$. Superscripts
 $ (\cdot)^{H}$ and $\mathrm{Tr}(\cdot)$ denote respectively the
Hermitian transpose and  trace  of a matrix. $A(m,n)$ represents the
$(m,n)$th entry of the matrix $\bf A$.

\section{Signal Model for CS-based  MIMO Radar}\label{sig_model}
Let us consider the same setting as in \cite{Yu:09_tsp}. In
particular, assume $K$ point targets and colocated transmit
(TX)/receive (RX) antennas that are randomly distributed over a
small area.
 The $k$-th target is at azimuth angle $\theta_k$ and moves
with constant radial speed $v_k$. Let $(r^t_{i},
\alpha^t_{i})$/$(r^r_{i}, \alpha^r_{i})$   denote the location of
the $i$-th TX/RX node   in polar coordinates. The number of TX and
RX nodes is denoted by $M_t$ and $N_r$, respectively. Let $d_k(t)$
denote the range of the $k$-th target at time $t$.
 Under the far-field assumption, i.e.,
 $d_{k}(t) \gg r^{t/r}_{i}$, the distance between the $i$th transmit/receive
node  and the $k$-th target
 $d^t_{ik}$/$d^r_{ik}$ can be approximated as
\begin{eqnarray}
d^{t/r}_{ik}(t) \approx d_k(t)- \eta_i^{t/r}(\theta_k)
=d_k(0)-\eta_{i}^{t/r}(\theta_k)-v_kt
\end{eqnarray}
where
$\eta_{i}^{t/r}(\theta_k)=r^{t/r}_{i}\cos(\theta_k-\alpha^{t/r}_{i})$.

Assuming that  there is no clutter, each TX node transmits  periodic
pulses of duration $T_p$ and pulse repetition interval (PRI)
 $T$. The  target return from the $k$-th target arriving at the $l$-th antenna during the
$m$-th pulse is
\begin{align} \label{rec_sig_1}
y^k_{lm}(t)&= \sum_{i=1}^{M_t}\beta_k
x_i(t-(d^{t}_{ik}(t)+d^{r}_{lk}(t))/c)
\exp({j{2\pi}f(t-(d^t_{ik}(t)+d^{r}_{lk}(t))/{c}) })
\end{align}
where $c$, $f$  and $\beta_{k}$  denotes  the speed of light, the
carrier frequency, and the reflection coefficient of the $k$-th
target, respectively;  $x_i(t)$ represents  the transmit waveform of
the $i$-th node. The demodulated baseband signal corresponding to a
single target can be approximated by
\begin{align} \label{rec_sig_1}
y^k_{lm}(t)& \approx \sum_{i=1}^{M_t}\beta_k x_i(t-2d_k(0)/c)
\exp(-{j{2\pi}f(d^t_{ik}(t)+d^{r}_{lk}(t))/{c} }).
\end{align}
In the above equation,  the time delays in  the received
waveforms due to the $k$-th target are all the same  and equal to
$2d_k(0)/c$. This approximation is enabled by   the assumption of
narrowband transmit waveforms, slowly moving targets and colocated
nodes.  The fact that the targets can be in different range bins
implies that the delays corresponding to different targets will be
different. Therefore, sampling the received signal and ensuring that
all target returns fall in the sampling window  would require time
delay estimation. However, in a CS scenario, there are not enough
data to obtain such estimates using traditional methods, e.g., the
matched filtering  method (MFM). In the following, we will extract
the range and angle information simultaneously using the CS approach
without assuming availability of time delay estimates.

The  compressed samples collected by the $l$-th antenna during the
$m$-th pulse can be expressed as
\begin{align}\label{rec_sig}
{{\bf r}}_{lm}&= \sum_{k=1}^{K}\beta_ke^{j2\pi {p}_{lmk}}{\bf
\Phi}_l{\bf D}(f_{k})\mathbf{C}_{\tau_k}{\bf X}{\bf
v}(\theta_k)+{\bf \Phi}_l{\bf n}_{lm}
\end{align}
where
\begin{enumerate}
\item ${p}_{lmk}=\frac{-2d_k(0) f}{c}+\frac{\eta_{l}^{r}(\theta_k)f}{c}+{f_{k}(m-1)T}$, where
$f_{k}=\frac{2v_k f}{c}$ is the Doppler shift induced by the $k$-th
target; $\mathrm{diag}\{{\bf X}^H{\bf X}\}=[1,\ldots,1]^T$; $lT_s,
l=0,\ldots,L-1$, represent the time within the pulse (fast time)
and thus the pulse duration is $T_p=LT_s$; 
\item $\mathbf{\Phi}_l$
 is the $M\times (L+\tilde{L})$ measurement matrix for the $l$-th receive
node where $\tilde{L}T_s$ is the maximum time delay and known in
advance. The measurement matrix has elements that  are independent
and identically distributed (i.i.d) Gaussian random variables;
\item ${\bf v}(\theta_k)=[e^{j\frac{2\pi
f}{c}\eta^t_{1}(\theta_k)},...,e^{j\frac{2\pi
f}{c}\eta^t_{M_t}(\theta_k)}]^T$
 and ${\bf D}(f_{k})={\rm
diag}\{[e^{j{2\pi}f_{k}0T_s},\ldots,e^{j{2\pi}f_{k}(L-1)T_s}]\}$;
\item $\tau_k=\lfloor\frac{2d_k(0)}{cT_s}\rfloor$ and ${\bf C}_{\tau_k}=[{\bf 0}_{L\times \tau_k},{\bf I}_{L},{\bf 0}_{L\times
(\tilde{L}-\tau_k)}]^T$. Here, we assume that the target returns
completely fall within the sampling window of length
$(L+\tilde{L})T_s$, and that $T_s$ is small enough so that the
rounding error in the delay is small, i.e., $x_i(t-\tau_k)\approx
x_i(t-\lfloor\frac{2d_k(0)}{cT_s}\rfloor)$.
\item ${\bf n}_{lm}$ is
the interference at the $l$-th receiver during the $m$-th pulse,
which includes a jammer's signal and thermal noise.
\end{enumerate}

  Let us  discretize the angle, speed and range
space on a fine grid, i.e., respectively, $[\tilde{a}_1,\ldots,\tilde{a}_{N_a}]$,
$[\tilde{b}_1,\ldots,\tilde{b}_{N_b}]$ and
$[\tilde{c}_1,\ldots,\tilde{c}_{N_c}]$. Let the grid points be arranged first angle-wise,
then
range-wise,  and finally speed-wise to yield the grid points $(a_n,b_n,c_n),
 n=1,...,N_a N_b N_c$.
 Through this ordering,   the grid point $(\tilde{a}_{n_a},\tilde{b}_{n_b},\tilde{c}_{n_c})$ is mapped to point
   $(a_n,b_n,c_n)$  with
$n=(n_b-1)n_an_c+(n_c-1)n_a+n_a$.
We assume that the discretization step is small enough so that each target
falls on some angle-speed-range
 grid point. Then (\ref{rec_sig})   can be
rewritten as
\begin{align}\label{received signal}
{{\bf r}}_{lm}&={\bf \Phi}_l\left( \sum_{n=1}^{N}s_ne^{j2\pi
{q}_{lmn}}{\bf
D}\left(\frac{2b_nf}{c}\right)\mathbf{C}_{\lfloor\frac{2c_n}{cT_s}\rfloor}{\bf
X}{\bf v}(a_n)+{\bf n}_{lm}\right)
\end{align}
where  $
 s_n = \left\{
\begin{array}{ll}
\mbox{ reflection coefficient of the target},  &  \text{if there is a target at}\ (a_n,b_n,c_n) \\
0,  & \text{if there is no target at} \  (a_n,b_n,c_n)
\end{array} \right.,$
$N=N_aN_bN_c$, and
\begin{align}
q_{lmn}=\frac{-2c_n
f}{c}+\frac{\eta_{l}^{r}(a_n)f}{c}+\frac{2b_nf(m-1)T}{c}.
\end{align}
 In  matrix form we have
 \begin{equation}
  {{\mathbf
r}}_{lm}=\mathbf{\Theta}_{lm}{\mathbf{s}}+{\bf \Phi}_l{\bf n}_{lm}
\end{equation}
 where  ${\bf s}=[s_1,...,s_N]^T$ and
\begin{align}\label{sensing_matrix}
\mathbf{\Theta}_{lm}={\bf \Phi}_l\underbrace{[e^{j2\pi q_{lm1}}{\bf
D}(2b_1f/c)\mathbf{C}_{\lfloor\frac{2c_1}{cT_s}\rfloor}{{\bf X}}{\bf
v}(a_1),\ldots,e^{j2\pi q_{lmN}}{\bf
D}(2b_Nf/c)\mathbf{C}_{\lfloor\frac{2c_N}{cT_s}\rfloor}{{\bf X}}{\bf
v}(a_N)]}_{\mathbf{\Psi}_{lm}}.
\end{align}
According to the CS formulation, $\mathbf{\Theta}_{lm}$ is the sensing matrix
 and $\mathbf{\Psi}_{lm}$ is the  basis matrix.

If the number of targets is small as compared to $N$,  the positions of
the targets are sparse in the angle-speed-range space, i.e.,
$\mathbf{s}$ is a sparse vector. The locations of the non-zero
elements of ${\bf s}$ provide information on target angle, speed and
range.

All the receive nodes forward their compressed measurements to a
fusion center. We assume that the fusion center has the ability to  separate
the data of different nodes from each other. This can be done,  for instance,  if the
nodes send their data over different carriers.
The fusion center
 combines the compressively sampled signals due to $N_p$ pulses
obtained at $N_r$ receive nodes to form the vector  ${{\bf r}}$. It holds that
\begin{eqnarray}\label{cs}
{{\bf r}}& \buildrel \triangle \over = &[{{\bf
r}}^T_{11},\ldots,{{\bf r}}^T_{1N_p},\ldots,{{\bf
r}}^T_{N_rN_p}]^T=\mathbf{\Theta}\mathbf{s}+{\bf n}
\end{eqnarray}
where $\mathbf{\Theta}=[(\mathbf{\Theta}_{11})^T,\ldots,(
\mathbf{\Theta}_{1N_p})^T,\ldots,(\mathbf{\Theta}_{N_rN_p})^T]^T$
and
 ${\bf n}=[({\bf \Phi}_1{\bf n}_{11})^T,\ldots,
({\bf \Phi}_1{\bf n}_{1N_p})^T,\ldots, ({\bf \Phi}_{N_r}{\bf
n}_{N_rN_p})^T]^T$.

Subsequently, using the predefined measurement matrices,
${\mathbf{\Phi}}_l,\ l=1,...,N_r$, based on the discretization of
the angle-speed-range space, and also based on knowledge of the
waveform matrix ${\bf X}$, the fusion center  recovers $\mathbf{s}$
by applying the Dantzig selector \cite{Candes:07} to  (\ref{cs}) as
\begin{eqnarray}\label{Dantzig}
\hat{{\bf s}}= \min\|{\bf s}\|_1\ \ \ s.t.\
\|{\mathbf{\Theta}}^H({\bf r}-\mathbf{\Theta}{\bf
s})\|_{\infty}<\mu.
\end{eqnarray}
According to \cite{Candes:07},  the sparse vector ${\bf s}$ can be
recovered  with very high probability if $\mu=(1+t^{-1})\sqrt{2\log
N\tilde \sigma^2}\sigma_{max}$, where $t$ is a positive scalar,
$\sigma_{max}$ is the maximum norm of columns in the sensing matrix
$\Theta$, and $\tilde {\sigma^2}$ is the variance of the
interference in (\ref{cs}).  A numerical method to determine the
value of $\mu$ is  described in \cite{Candes:07}.

\section{Introducing Step Frequency to CS-MIMO radar}\label{mod_sf}

Let us consider a MIMO radar system in which the carrier frequency
of the $m$-th pulse equals
\begin{align}
f_m=f+\Delta f_m
\end{align}
where $f$ is the center carrier frequency and $\Delta f_m$ denotes
the frequency step,  $m=1,\ldots, N_p$.

The  baseband samples collected by the $l$-th antenna during the
$m$-th pulse can be expressed as
\begin{align}\label{rec_sig_sf1}
\tilde{{\bf r}}_{lm}&= {\bf \Phi}_l\sum_{k=1}^{K}\beta_ke^{j2\pi
\tilde{p}_{lmk}}{\bf D}(f_{mk})\mathbf{C}_{\tau_k}{\bf X}{\bf
v}_m(\theta_k)+{\bf \Phi}_l{\bf n}_{lm}
\end{align}
 where
\begin{align}
\ f_{mk}&=\frac{2v_k f_m}{c}, {\bf v}_m(\theta_k)=[e^{j\frac{2\pi
f_m}{c}\eta^t_{1}(\theta_k)},...,e^{j\frac{2\pi
f_m}{c}\eta^t_{M_t}(\theta_k)}]^T\nonumber\\
\mathrm{and}\ \tilde{p}_{lmk}&=\frac{-2d_k(0)
f_m}{c}+\frac{\eta_{l}^{r}(\theta_k)f_m}{c}+{f_{mk}(m-1)T}.
\end{align}
Then, based on discrete grid points of the angle-speed-range space,
(\ref{rec_sig_sf1}) can be rewritten as
\begin{eqnarray}
\tilde{{\bf r}}_{lm}&=& {\bf \Phi}_l\tilde{{\bf\Psi}}_{lm}{\bf
s}+{\bf \Phi}_l{\bf
n}_{lm} \nonumber \\
&=&\tilde{\mathbf{\Theta}}_{lm}{\bf s}+{\bf \Phi}_l{\bf n}_{lm}
\end{eqnarray}
where
\begin{align}
\tilde{\mathbf{\Psi}}_{lm}&=[e^{j2\pi \tilde{q}_{lm1}}{\bf
D}(2b_1f_m/c)\mathbf{C}_{\lfloor\frac{2c_1}{cT_s}\rfloor}{{\bf
X}}{\bf v}_m(a_1),\ldots,e^{j2\pi \tilde{q}_{lmN}}{\bf
D}(2b_Nf_m/c)\mathbf{C}_{\lfloor\frac{2c_N}{cT_s}\rfloor}{{\bf
X}}{\bf v}_m(a_N)],\nonumber \\
 \tilde{q}_{lmn}&=\frac{-2c_n
f_m}{c}+\frac{\eta_{l}^{r}(a_n)f_m}{c}+\frac{2b_nf_m(m-1)T}{c},\nonumber\\
\mathrm{ and}\ \tilde{{\bf \Theta}}_{lm}&={\bf \Phi}_l\tilde{{\bf
\Psi}}_{lm}.
\end{align}

At the fusion center,
 the compressively sampled signals due to $N_p$ pulses
obtained at $N_r$ receive nodes are stacked as
\begin{eqnarray}
\tilde{{\bf r}}& \buildrel \triangle \over =
\tilde{\mathbf{\Theta}}\mathbf{s}+{\bf n}
\end{eqnarray}
where \begin{align}\label{sensing_matrix_all}
\tilde{\mathbf{\Theta}}=[(\tilde{\mathbf{\Theta}}_{11})^T,\ldots,(
\tilde{\mathbf{\Theta}}_{1N_p})^T,\ldots,(\tilde{\mathbf{\Theta}}_{N_rN_p})^T]^T.
\end{align}
Recovery of $\mathbf{s}$ is performed as in (\ref{Dantzig}) where $\mathbf{\Theta}$ is replaced with $\tilde{\mathbf{\Theta}}$.

In the remainder of the paper, we make the two assumptions:

\begin{itemize}
\item
(A1) The targets are slowly moving. Therefore, the Doppler shift
within a pulse can be ignored, i.e., $
f_m(2T_s(L+\tilde{L}-1)b_n)/c\approx 0,\ n=1,\ldots, N$.

\item (A2) The radar waveforms are independent across transmit nodes and thus
$\int_{t=0}^{T}x_i(t)x^*_{i'}(t+\tau)dt, i\neq i'$ is negligible as
compared to $\int_{t=0}^{T}x_i(t)x^*_{i}(t+\tau) dt$.
\end{itemize}

\subsection{Range resolution}\label{sec_range_resolution}

In this subsection we study the relationship between range
resolution and the ambiguity function.  For the conventional  radar
systems that uses a matched filter to extract  target information,
the ambiguity function (AF) characterizes the response to a point
target and determines resolution.  Let us assume that there is a
target at $(\theta,d,v)$. The matched filter looking for a target at
$(\theta',d',v')$  yields
\begin{align} \label{AF_CSSF}
\chi(\Delta d,\Delta v,
\theta,\theta')=\sum_{l=1}^{N_r}\sum_{i,i'=1}^{M_t}\sum_{m=1}^{N_p}\chi_{i,i',m}(\Delta
d,\Delta v)e^{j2\pi
f_m\frac{\eta_i^t(\theta)+\eta_l^r(\theta)-\eta_{i'}^t(\theta')-\eta_l^r(\theta')-2\Delta
d}{c}}
\end{align}
where $\Delta d=d-d'$, $\Delta v=v-v'$ and
\begin{align}
\chi_{i,i',m}(\Delta d,\Delta v)\triangleq
\int_{t}x_i(t)x^*_{i'}(t+2\Delta d/c)e^{j2\pi f_m\frac{2\Delta
v}{c}t}dt.
\end{align}
Equation (\ref{AF_CSSF}) is the AF for SF MIMO radar, where SF MIMO
radar  refers to   MIMO radar that uses the SF technique. Unlike the
AF for MIMO radar \cite{Chen:082},   the carrier frequency is
varying between pulses in  (\ref{AF_CSSF}).

 To investigate the range resolution let us
set $\Delta v=0$ and $\theta=\theta'$. Then, the AF becomes
\begin{align} \label{AF_range}
\chi(\Delta d,0,
\theta,\theta)&=N_r\sum_{i,i'=1}^{M_t}\sum_{m=1}^{N_p}\chi_{i,i',m}(\Delta
d,0)e^{j2\pi f_m\frac{\eta_i^t(\theta)-\eta_{i'}^t(\theta)-2\Delta
d}{c}}\nonumber\\
&=N_r\underbrace{\sum_{m=1}^{N_p}e^{j2\pi f_m(-2\Delta
d/c)}}_{\chi_1(\Delta
d)}\underbrace{\sum_{i=i'}\int_{t}x_i(t)x^*_{i'}(t+2\Delta
d/c)}_{\chi_2(\Delta
d)}dt+N_r\underbrace{\sum_{m=1}^{N_p}\sum_{i\neq i'}e^{j2\pi
f_m\frac{\eta_i^t(\theta)-\eta_{i'}^t(\theta)-2\Delta
d}{c}}\int_{t}x_i(t)x^*_{i'}(t+2\Delta
d/c)}_{\Delta \chi(\Delta d)}dt\nonumber\\
\end{align}

Due to (A2), the term $\Delta \chi(\Delta d)$ is negligible as
compared to the product $\chi_1(\Delta d)\chi_2(\Delta d)$ in
(\ref{AF_range}). One can see
 that $\chi_1(\Delta d)$ and $\chi_2(\Delta
d)$ are respectively the AF of SF single-input single-output (SISO)
radar
 and  MIMO radar, both for $\Delta v=0$ and $\theta=\theta'$. It can
 seen from (\ref{AF_range}) that
 a colocated MIMO radar has no gain on range resolution as
 compared to a SISO radar, i.e.,
 the range resolution of MFSF MIMO radar  is  at least equal to the
best between the range resolution of
 SF SISO radar and  SISO radar, where MFSF MIMO radar  refers to matched filter based  MIMO radar that uses the SF technique.

In \cite{Song:10},
in a study of CS-based  SISO
 radar,
 it was observed that the maximum value of  the correlation of two different columns of
the basis matrix  is equal to the second largest value of the
discrete AF surface.  The recovery performance of CS approaches,
however, is directly related to the column correlation of   the
sensing matrix rather than  the basis matrix. Unlike \cite{Song:10},
we next study the relation of the AF and the column correlation of
the sensing matrix
 for the proposed CSSF MIMO radar. This analysis will provide a clue   for comparing the resolution of CS and matched filter in the context of SF MIMO radar, i.e., CSSF MIMO radar and
 MFSF MIMO radar.

 On letting ${\bf p}_k$ denote the
column of the sensing matrix $\tilde{\mathbf{\Theta}}$
corresponding to the $k$-th grid point in the angle-speed-range
space, we have
\begin{align}\label{corr}
<{\bf p}_k,{\bf
p}_{k'}>&=\sum_{l=1}^{N_r}\sum_{m=1}^{N_p}e^{j2\pi(\tilde{q}_{lmk}-\tilde{q}_{lmk'})}
{\bf v}_m^H(a_{k'}){\bf
    X}^H{\bf C}_{\lfloor \frac{2c_{k'}}{cT_s}\rfloor}^H{\bf D}^H\left(\frac{2b_{k'}f_m}{c}\right)\underbrace{{\mathbf{\Phi}}_l^H{\mathbf{\Phi}}_l}_{{\bf A}}\underbrace{{\bf D}\left(\frac{2b_{k}f_m}{c}\right){\bf C}_{\lfloor \frac{2c_{k}}{cT_s}\rfloor}{\bf
    X}{\bf v}_m(a_{k})}_{\mathbf{g}_k} \nonumber\\
    &=\sum_{l=1}^{N_r}\sum_{m=1}^{N_p}\sum_{p,q=1}^{L+\tilde{L}}e^{j2\pi(\tilde{q}_{lmk}-\tilde{q}_{lmk'})}{g}^*_{k'}(p){g}_k(q){
    A}(p,q)
    \nonumber\\
&=\sum_{l=1}^{N_r}\sum_{m=1}^{N_p}\sum_{p,q=1}^{L+\tilde{L}}\sum_{i,i'=1}^{M_t}{
    A}(p,q)e^{j2\pi
f_m(\eta_i^t(a_k)+\eta_l^r(a_k)-\eta_{i'}^t(a_{k'})-\eta_l^r(a_{k'})-2\Delta
d_{kk'}+2\Delta
v_{kk'}(m-1)T+2T_s(b_k(q-1)-b_{k'}(p-1)))/c}\nonumber\\
&\cdot
x_i\left((q-1)T_s-\frac{2c_k}{c}\right)x^*_{i'}\left((p-1)T_s-\frac{2c_{k'}}{c}\right)
\end{align}
where $\Delta d_{kk'}=c_k-c_{k'}$ and $\Delta v_{kk'}=b_k-b_{k'}$.
For simplicity, in the above we assumed that the receive nodes use
the same measurement matrix; thus the index $l$ was dropped in ${\bf
A}$.

Taking the elements of the measurement matrix ${\bf \Phi}$ to be
independent and Gaussian
 $\mathcal{N}(0,
\frac{1}{L+\tilde{L}})$,  the expectation of $<{\bf p}_k,{\bf
p}_{k'}>$ with respect to the elements of ${\bf \Phi}$ equals
\begin{align}\label{corr_expect}
E\{<{\bf p}_k,{\bf
p}_{k'}>\}&=\frac{M}{L+\tilde{L}}\sum_{l=1}^{N_r}\sum_{m=1}^{N_p}\sum_{i,i'=1}^{M_t}e^{j2\pi
f_m(\eta_i^t(a_k)+\eta_l^r(a_k)-\eta_{i'}^t(a_{k'})-\eta_l^r(a_{k'})-2\Delta
d_{kk'}+2\Delta
v_{kk'}(m-1)T)/c}\nonumber\\
&\cdot
\sum_{p=1}^{L+\tilde{L}}x_i\left((p-1)T_s-\frac{2c_k}{c}\right)x^*_{i'}\left((p-1)T_s-\frac{2c_{k'}}{c}\right)e^{j2\pi
f_m(2T_s(p-1)\Delta v_{kk'})/c}\nonumber\\
&\propto \chi(\Delta d_{kk'},\Delta v_{kk'}, a_k,a_{k'}).
\end{align}
One can see from the above equation that the expectation of the
column correlation of the sensing matrix is proportional to the
discrete AF. To focus on the range resolution we set $a_k=a_{k'}$
and $\Delta v_{kk'}=0$.
 Essentially,  the range
resolution of MFSF MIMO  radar corresponds to
the smallest range difference between two targets, $\Delta d_{kk'}$,
that sets the AF to zero.
 Based on the UUP in \cite{Candes:062}, however, the coherence of the sensing matrix does not have to be zero
 for exact recovery; a  small level of coherence is good enough.
 Therefore, CS-based radar
 systems have the potential to improve
  range resolution. This possibility will be confirmed via simulations in Section \ref{simulation} (see Fig. \ref{com_range_resolution}).

\subsection{The effect of signal bandwidth on CSSF-MIMO radar }\label{mod_sf}
In an LSFR system, the carrier frequency increases by a constant
step between pulses, i.e., $\Delta f_m=(m-1)\Delta f$. This type of
SF radar can be efficiently implemented using the Inverse Discrete
Fourier Transform (IDFT) \cite{Gill}; however, it suffers from range
ambiguity if the distance between a target and receive nodes exceeds
the value $R_u=\frac{cT}{2}$. The range ambiguity can be removed by
randomly choosing the step frequency within a fixed bandwidth at the
expense of increased sidelobe as compared to the LSFR
\cite{Axelsson:07}. In this section, we investigate the effect of
the number of pulses $N_p$ (or equivalently, the  bandwidth) on
range resolution for  two types of CSSF MIMO radar, i.e., LSFR and
RSFR,  in terms of the coherence of the sensing matrix (see
(\ref{coherence})). Consistent with \cite{Axelsson:07}, which
discussed convectional radar systems using  the MFM, we find that
the RSFR requires more pulses than LSFR to achieve the same range
resolution for CS-based MIMO radar.

Since an increase in the number of receive nodes does not improve the
range resolution, for simplicity we consider  one receive node only.
  The correlation of columns ${\bf p}_k$ and ${\bf
p}_{k'}$ for $a_k=a_{k'}$ and $b_k=b_{k'}$ equals
\begin{align}\label{corr_sf}
p_{kk'}&=|<{\bf p}_k,{\bf p}_{k'}>|
    =\left|\sum_{m=1}^{N_p}\sum_{p,q=1}^{L+\tilde{L}}e^{j2\pi f_m(-2\Delta
d_{kk'})/c}{g}^*_{k'}(p){g}_k(q){
    A}(p,q)\right| \nonumber\\
    &=
\left|\sum_{m=1}^{N_p}e^{j2\pi f_m(-2\Delta
d_{kk'})/c}\sum_{p,q=1}^{L+\tilde{L}}{
    A}(p,q)e^{j2\pi f_m(2T_sb_k(q-p))/c}
\left(\sum_{i=1}^{M_t}Q_{kk'}(m,p,q,i,i)+\sum_{i\neq
i'}^{M_t}Q_{kk'}(m,p,q,i,i')\right)\right|
\end{align}
where
\begin{align}
Q_{kk'}(m,p,q,i,i')=e^{j2\pi
f_m(\eta_i^t(a_k)-\eta_{i'}^t(a_{k}))/c}
x_i\left((q-1)T_s-\frac{2c_k}{c}\right)x^*_{i'}\left((p-1)T_s-\frac{2c_{k'}}{c}\right).
\end{align}

Due to (A1) and the discretized version of (A2),  we can ignore the Doppler
shift within a pulse  and the second term $\sum_{i\neq
i'}^{M_t}Q_{kk'}(m,p,q,i,i')$ in (\ref{corr_sf}). Therefore,
(\ref{corr_sf}) becomes
\begin{align}\label{corr_sf1}
p_{kk'}
    &\approx
\left|\sum_{m=1}^{N_p}e^{j2\pi f_m(-2\Delta
d_{kk'})/c}\sum_{p,q=1}^{L+\tilde{L}}{
    A}(p,q)
\sum_{i=1}^{M_t}x_i\left((q-1)T_s-\frac{2c_k}{c}\right)x^*_{i}\left((p-1)T_s-\frac{2c_{k'}}{c}\right)\right|.
\end{align}
Eq. (\ref{corr_sf1}) can be rewritten as
\begin{align}
&p_{kk'}\approx \left\{
\begin{array}{ll}
{N_p}\rho_{kk}& k=k'\\
\underbrace{\left|\sum_{m=1}^{N_p}e^{j\alpha_{kk'} (f+\Delta
f_m)}\right|}_{h(\mathbf{\Delta f})}\rho_{kk'}& k\neq k'
\end{array} \right.\
\end{align}
where  $\mathbf{\Delta} f=[\Delta f_1,\ldots,\Delta f_{N_p}]$,
\begin{align}
\rho_{kk'}&=\left|\sum_{p,q=1}^{L+\tilde{L}}\sum_{i=1}^{M_t}{
    A}(p,q)
x_i((q-1)T_s-\frac{2c_k}{c})x^*_{i}((p-1)T_s-\frac{2c_{k'}}{c})\right|\nonumber\\
\mathrm{and}\  \alpha_{kk'}&=-\frac{4\pi\Delta
d_{kk'}}{c}.\label{a_kk}
\end{align}

Then, the coherence of the sensing matrix $\tilde{\bf \Theta}$
corresponding to columns ${\bf p}_k$ and ${\bf p}_{k'}$ can be
written as
\begin{align}\label{coherence}
\mu_{kk'}(\tilde{\bf
\Theta})=\frac{p_{kk'}}{\sqrt{p_{kk}p_{k'k'}}}\approx
\frac{h(\mathbf{\Delta
f})\rho_{kk'}}{N_p\sqrt{\rho_{kk}\rho_{k'k'}}}.
\end{align}

\subsubsection{Linear step frequency}
If the carrier frequency increases by a constant step $\Delta f$
between adjacent pulses, i.e.,  $\Delta f_m=(m-1)\Delta f$, then
\begin{align}\label{constant_step}
\mu_{kk'}(\tilde{\bf \Theta})&\approx \frac{|1-e^{j\alpha_{kk'}
\Delta fN_p}|\rho_{kk'}}{|1-e^{j\alpha_{kk'} \Delta
f}|N_p\sqrt{\rho_{kk}\rho_{k'k'}}}\propto
\frac{|\sin(\frac{1}{2}\alpha_{kk'} \Delta fN_p)|}{N_p}.
\end{align}
It can be easily seen that an increase in $N_p$ tends to reduce the
coherence and thus improves the range resolution.

  Let $\alpha_{kk'ii'}^{pq}$ denote the
travel-time difference between the signals sent from  the transmit
node $i$ to the target located at the $k$th grid point at  time
instant $pT_s$, and from  the transmit node $i'$ to the target
located at the $k'$th grid point at time instant $qT_s$. It holds
that
\begin{align}\label{a_kkii}
\alpha_{kk'ii'}^{pq}=(-2\Delta
d_{kk'}+2T_sb_k(q-p)+\eta_i^t(a_k)-\eta_{i'}^t(a_{k}))/c.
\end{align}

Regarding the approximation error, the term discarded in
(\ref{corr_sf1}) is
\begin{align}
\tilde{p}_{kk'}
   & =
\sum_{m=1}^{N_p}e^{j2\pi f_m(-2\Delta
d_{kk'})/c}\sum_{p,q=1}^{L+\tilde{L}}{
    A}(p,q)e^{j2\pi f_m(2T_sb_k(q-p))/c}
\sum_{i\neq i'}^{M_t}Q_{kk'}(m,p,q,i,i')\nonumber\\
&=\sum_{p,q=1}^{L+\tilde{L}}\frac{1-e^{j2\pi N_p\Delta
f\alpha_{kk'ii'}^{pq}}}{1-e^{j 2\pi\Delta
f\alpha_{kk'ii'}^{pq}}}e^{j 2\pi f\alpha_{kk'ii'}^{pq}}{
    A}(p,q)
\sum_{i\neq i'}^{M_t}
x_i\left((q-1)T_s-\frac{2c_k}{c}\right)x^*_{i'}\left((p-1)T_s-\frac{2c_{k'}}{c}\right).
\end{align}
The amplitude of $\frac{1-e^{j2\pi N_p\Delta
f\alpha_{kk'ii'}^{pq}}}{1-e^{j 2\pi \Delta
f\alpha_{kk'ii'}^{pq}}}e^{j 2\pi f\alpha_{kk'ii'}^{pq}}$ is bounded
by $N_p$. For  independent waveforms, the approximation error
$\tilde{p}_{kk'}$ in (\ref{corr_sf}) is always negligible as
compared to ${p}_{kk'}$.

Let $\mu_t$ denote the maximum coherence of $\tilde{\bf \Theta}$
that guarantees  exact recovery of the sparse vector with high
probability via the Dantzig selector. The minimum number of pulses
required to achieve a certain resolution can be obtained by solving
\begin{align}
N_p^*&=\min\  N_p\nonumber\\
 &s.t.\ \frac{|1-e^{j\alpha_{kk'} \Delta
fN_p}|\rho_{kk'}}{|1-e^{j\alpha_{kk'} \Delta f}|N_p\sqrt{\rho_{kk}\rho_{k'k'}}}\leq \mu_t,\nonumber\\
& k,k'=1,\ldots,N\ \mathrm{and}\ k\neq k'.
\end{align}

The above problem is easy to solve, for example by trying different
values for $N_p$; however, it requires a value for $\mu_t$. In
\cite{Strohmer:Asilomar09},  a rough estimate of $\mu_t$ in the
presence of mild interference was offered.  In general, $\mu_t$ must
be determined experimentally.

\subsubsection{Random step frequency}
Assuming that the frequency steps over pulses are i.i.d uniform
random variables, i.e., $\Delta f_m \sim U(0,2b)$,  the expectation
of square coherence over $\Delta f_m$ is given by
\begin{align}\label{random_step}
E\{\mu^2_{kk'}(\tilde{\bf
\Theta})\}&=E\left\{\frac{\left|\sum_{m=1}^{N_p}e^{j\alpha_{kk'}(f+\Delta
f_m)}\rho_{kk'}\right|^2}{N_p^2{\rho_{kk}\rho_{k'k'}}}\right\}\nonumber\\
&=\frac{\rho_{kk'}^2}{{\rho_{kk}\rho_{k'k'}}}\left(
\frac{1}{N_p}+\frac{N^2_p-N_p}{N^2_p}
\frac{\sin^2(\alpha_{kk'}b)}{\alpha_{kk'}^2b^2}\right).
\end{align}
 For a fair comparison, we set
LSFR and RSFR to cover the same frequency band, i.e., set $b$ equal
to $\Delta f(N_p-1)/2$. Then (\ref{random_step}) can be rewritten
as
\begin{eqnarray}\label{random_step1}
E\{\mu^2_{kk'}(\tilde{\bf
\Theta})\}&=\frac{\rho^2_{kk'}}{{\rho_{kk}\rho_{k'k'}}N_p}\left(1+
\frac{4\sin^2(\frac{1}{2}(N_p-1)\alpha_{kk'}\Delta
f)}{(N_p-1)\alpha_{kk'}^2\Delta f^2}\right)\nonumber\\
&=\frac{\rho^2_{kk'}}{{\rho_{kk}\rho_{k'k'}}N_p}\left(1+
\frac{\sin^2(\frac{1}{2}(N_p-1)\alpha_{kk'}\Delta
f)}{(N_p-1)(2\pi\Delta f \Delta d_{kk'}/c)^2}\right).
\end{eqnarray}
As  the term $(2\pi\Delta f\Delta d_{kk'}/c)^2$ increases, the
expected value of the squared coherence  becomes approximately equal
to $1/N_p$. This holds when the product of radian frequency step
$2\pi\Delta f$ and  the range spacing of grid points $\Delta
d_{kk'}$ is  comparable to the speed of light $c$.


Since the coherence of the sensing matrix for  RSFR cannot be
 obtained directly, we  instead compare the squared coherence of the sensing
matrix for LSFR and RSFR. For large $N_p$, we find from
(\ref{constant_step}) and (\ref{random_step1}) that the squared
coherence for LSFR and RSFR decreases  inverse proportionally to
$N^2_p$ and $N_p$, respectively. This implies that more pulses are
required by RSFR to achieve the desired performance with all other
parameters, i.e., $M_t$, $N_r$ and $M$,  being equal.

Before ending  this section, we note that the expectation of the approximation error in (\ref{corr_sf})
can be represented by
\begin{align}
\tilde{p}_{kk'}
   & =
\sum_{p,q=1}^{L+\tilde{L}}e^{j 2\pi \alpha^{pq}_{kk'ii'}
(N_p-1)\Delta
f/2+f}\frac{N_p}{N_p-1}\frac{2\sin(\pi\alpha^{pq}_{kk'ii'}(N_p-1)\Delta
f)}{{\alpha^{pq}_{kk'ii'}2\pi\Delta f}\bf
    }A(p,q)
\sum_{i\neq i'}^{M_t}
x_i\left((q-1)T_s-\frac{2c_k}{c}\right)x^*_{i'}\left((p-1)T_s-\frac{2c_{k'}}{c}\right)
\end{align}
where one can see that a decrease in the product of
$\alpha^{pq}_{kk'ii'}$ (seen in (\ref{a_kkii})) and the radian
frequency step, $2\pi\Delta f$,   increases both the approximation
error and the squared coherence. Given $\alpha^{pq}_{kk'ii'}$,  an
increase in  $\Delta f$ would reduce the approximation error
$\tilde{p}_{kk'}$. However, this would increase the bandwidth
required by RSFR.

\section{Decoupled estimation of angle,
velocity and range with reduced complexity}\label{decoupled_scheme}

Solving the $\ell_1$ minimization problem of (\ref{Dantzig})
requires polynomial time in the dimension of $\mathbf{s}$. For the
discretization discussed in Section II,  the joint estimation of
angle, velocity and range requires complexity of
 $\mathcal{O}((N_aN_bN_c)^3)$ \cite{Candes:05}\cite{Candes:07}. For large values of $N_a$, $N_b$ and $N_c$, the
 computational cost of the CS approach would be prohibitive.
In the following, we propose a decoupled angle-velocity-range
estimation approach which reduces the search space and thus the
computational complexity.

The scheme needs some initial rough estimates of angle and range.
One way to obtain those estimates is to use the MFM, which requires
forwarding to the fusion center Nyquist sampled data from one pulse.
In the following, all $N_r$ nodes in the system sample all received
pulses in a compressive fashion, except $\tilde{N_r}$ nodes, which
sample the first received pulse at  the Nyquist rate and all
remaining pulses in a compressive fashion. Those Nyquist rate
samples will be used  to obtain coarse estimates of angle and range
via the MFM.

\begin{figure}[htbp]
  \centering
    \includegraphics[height=3.0in,width=4in,clip=true]{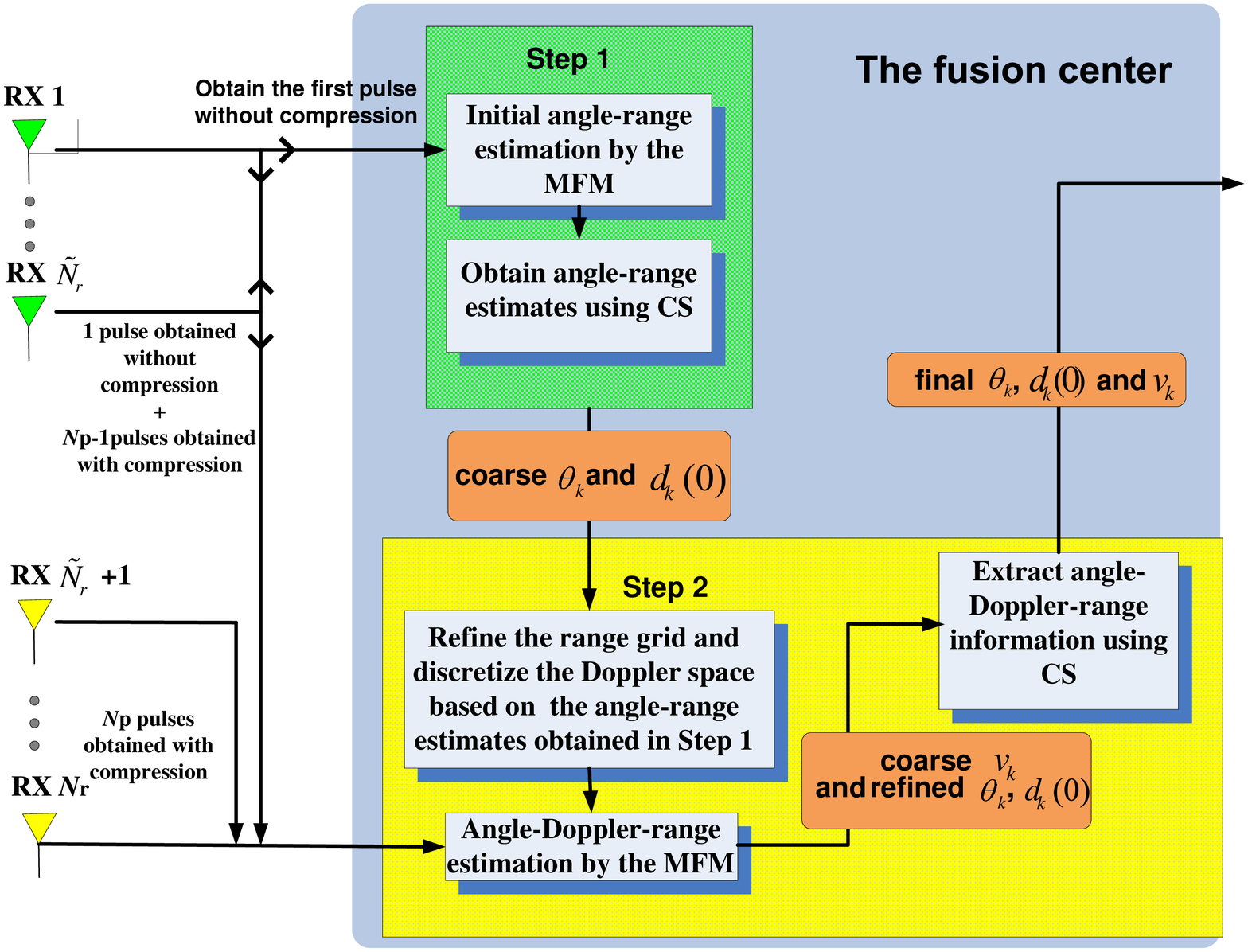}
    \caption{  \footnotesize{ Schematic diagram of the proposed decoupled scheme.
  }}\label{decoupling_scheme}
\end{figure}

The fusion center performs the following operations
 (also
see Fig. \ref{decoupling_scheme}).

\begin{enumerate}
\item  \emph{STEP1: Angle and range estimation}

This step uses the first pulse forwarded by each receive node. A
fine grid, $(a_{n_1}, c_{n_1}),\ldots,(a_{n_{K_1}},
 c_{n_{K_1}})$, is constructed around the MFM initial estimates. Then the sensing
 matrix  is constructed as
\begin{align}
\tilde{\mathbf{\Theta}}_1={\bf\Phi}_{l1}[e^{j2\pi f_1
(-2c_{n_1}+\eta_l^r(a_{n_1}))/c}\mathbf{C}_{\lfloor\frac{2c_{n_1}}{cT_s}\rfloor}{{\bf
X}}{\bf v}_m(a_{n_1}),\ldots,e^{j2\pi f_1
(-2c_{n_{K_1}}+\eta_l^r(a_{n_{K_1}}))/c}\mathbf{C}_{\lfloor\frac{2c_{n_{K_1}}}{cT_s}\rfloor}{{\bf
X}}{\bf v}_m(a_{n_{K_1}})]
\end{align}
where
\begin{align}
{{\bf\Phi}}_{lm}=\left\{
\begin{array}{rl}
{\bf I}_{L+\tilde{L}},& l=1,\ldots,\tilde{N_r},\ m=1\\
 \mathrm{the\ measurement\ matrix\ of\
size}\ M\times ({L+\tilde{L}}),& \mathrm{otherwise}
\end{array} \right.\ .
\end{align}
The received signals, $\tilde{\mathbf{r}}_{11},\ldots,
\tilde{\mathbf{r}}_{N_r1}$, are stacked in a vector, i.e.,
 \begin{align}\label{rec_step1}
\tilde{{\bf r}}_{1}=\tilde{\mathbf{\Theta}}_1{\bf s}+{\bf n}_1
 \end{align}
where $\tilde{{\bf r}}_1=[\tilde{\mathbf{r}}_{11}^T,\ldots,
\tilde{\mathbf{r}}^T_{N_r1}]^T$. By applying the Dantzig selector to
(\ref{rec_step1}),  new and refined angle-range information is
obtained.

Thanks to the initial estimates, the search area in the angle-range
plane is significantly reduced and thus the computational load of CS
is lightened. Due to the fact that only one pulse from each receive
node is used, the range resolution at this step is limited by
$\frac{c}{2B}$, where $B$ is the signal bandwidth. The obtained
range estimates will be refined in the next step in which the fusion
center will jointly process the entire pulse train. Also, due to
assumption (A1), Doppler information cannot be extracted at this
step.

\item  \emph{STEP 2: Range resolution improvement and Doppler estimation}

In this step the fusion center processes the entire pulse train
forwarded by each  receive node.  The range space around the
range estimates obtained in Step 1 is discretized into finer grid points. Based on a
discretization of the Doppler space, the refined range grid points
and the angle estimates obtained in Step 1, i.e., $(a_{m_1},
b_{m_1}, c_{m_1}),\ldots,(a_{m_{K_2}}, b_{m_{K_2}},
 c_{m_{K_2}})$,  the fusion center
 formulates a sensing matrix  and extracts angle-Doppler-range
 information in a CS fashion.

 To further reduce the complexity of CS reconstruction,
 the MFM can be applied before CS to provide  angle-Doppler-range estimates around which a finer grid can be
constructed and used by CS. In that case MFM would be applied based
on the  grid points  $(a_{m_1}, b_{m_1},
c_{m_1}),\ldots,(a_{m_{K_2}}, b_{m_{K_2}},
 c_{m_{K_2}})$.

For the case in which  there are stationary targets and moving targets, the angle estimation  can be further improved by taking into account Doppler information.

\end{enumerate}

Assuming that the MFM is used for initial estimation, the complexity
of two steps is respectively
$\mathcal{O}(N_aN_c(\tilde{N}_rL+(N_r-\tilde{N}_r)M)+K_1^3)$ and
$\mathcal{O}(K_2(\tilde{N}_r(L-M)+N_rN_pM)+K_3^3)$, where $K_3$ is
the number grid points used by CS at Step 2. Generally, it holds
that $K_1^3+K_3^3\ll N_aN_c(\tilde{N}_rL+(N_r-\tilde{N}_r)M)+
K_2(\tilde{N}_r(L-M)+N_rN_pM)$ for a small number of targets.
Therefore,
 the computational load is mostly
due to the  initial estimation.  As compared to the complexity of
the joint angle-Doppler-range CS approach, i.e.,
$\mathcal{O}((N_aN_bN_c)^3$, considerable computations can be saved
by using the proposed decoupled scheme for large values of $N_a,
N_b$ and $N_c$.

The
 computation savings, however, may be obtained at the expense of detection accuracy, unless the
initial estimates provided by the initial estimation method are
reliable. \emph{Reliable estimates} here refer to the initial
estimates whose distances to the true target locations are within
the resolution cell that is determined by the initial estimation.
Then all the targets can be captured based on the finer angle-range
grid points constructed around the reliable initial estimates.
 For the instance of the MFM, the
performance in providing good initial estimates depends on several
factors; (i) the signal-to-interference ratio (SIR), which can be
improved by employing more data; (ii)  angular, range or Doppler
resolution, which is improved by increasing $N_r$ or $N_p$; (iii)
the distance between the adjacent grid points. (In the worst case in
which the targets fall  midway between grid points, the targets may
fail to be captured by the closest grid points if the spacing of
adjacent grid points is too large. An empirical approach to select
grid spacing was discussed in \cite{Yu:09_tsp}. That approach is
also applicable to the MFM); and (iv) the threshold for hard
detection. A small threshold should be used in order to reduce the
miss probability. However, this implies
 that  more grid points need to be considered for the CS
approach following  the MFM as compared to a larger threshold. In
summary, the performance of the MFM can be improved at the expense
of more transmit power and increased complexity.

\section{Simulation Results}\label{simulation}
We consider a MIMO radar system with  transmit and  receive nodes
uniformly distributed on a disk of radius $10$m.   The  carrier
frequency is $f=5 GHz$.  Each transmit node uses 
orthogonal Hadamard waveforms of length $L=512$ and unit power. The
received signal is
 corrupted by  zero-mean Gaussian noise. The
signal-to-noise ratio (SNR)  is defined as the inverse  of the power
of thermal noise at a receive node. A jammer is located at angle $7
\textordmasculine$ and transmits an unknown Gaussian random
waveform. The targets are assumed to fall on the grid points.
Throughout this section, the CS approach uses   a  measurement
matrix with Gaussian entries.

\subsection{Range resolution of the CS-based SFR and conventional SFR}
\begin{figure}[htbp]
  \centering
    \includegraphics[height=3in,width=4in,clip=true]{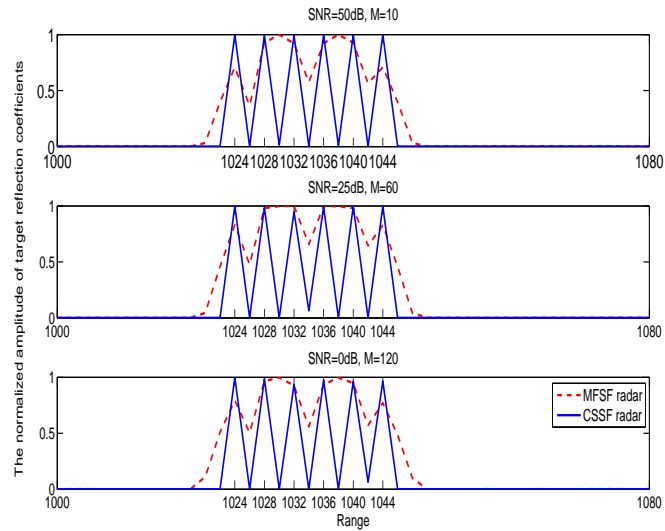}
\caption{  \footnotesize{Normalized amplitude estimates of target
reflection coefficients for the CSSF radar and MFSF radar  (one
realization for $M_t=N_r=1$ and $N_p=30$).
  }}\label{com_range_resolution}
\end{figure}

In this subsection we provide some simulation results to show the
superiority of  CSSF MIMO radar as compared to  MFSF MIMO radar in
terms of range resolution.  Figure \ref{com_range_resolution} shows
the normalized amplitude estimates of target reflection coefficients
for CSSF MIMO radar and MFSF MIMO radar  in one realization. Since
the multiple colocated antennas fail to improve range resolution, we
consider a single transmit and receive antenna  here for simplicity.
Let $M=10$, $N_p=30$ and the carrier frequencies be randomly
selected within the frequency band $[5, 5.029]GHz$. The CSSF radar
uses $10$ measurements per pulse while MFSF radar obtains $665$
measurements per pulse.  Various values of SNR are considered. The
spacing between two adjacent grid points is $2m$. There are six
targets at ranges $[1024,\ 1028,\ 1032,\ 1036,\ 1040,\ 1044]$m.
Figure \ref{com_range_resolution} shows  that the peaks
corresponding to all targets can be distinguished from each other
for the CSSF radar while for the MFSF radar some peaks are lost.
This verifies the observations of Section \ref{sec_range_resolution}
that  CSSF radar has the potential to achieve higher range
resolution than does MFSF radar.
\subsection{Range estimation for CSSF MIMO radar}
\begin{figure}[htbp]
  \centering
    \includegraphics[height=3in,width=4in,clip=true]{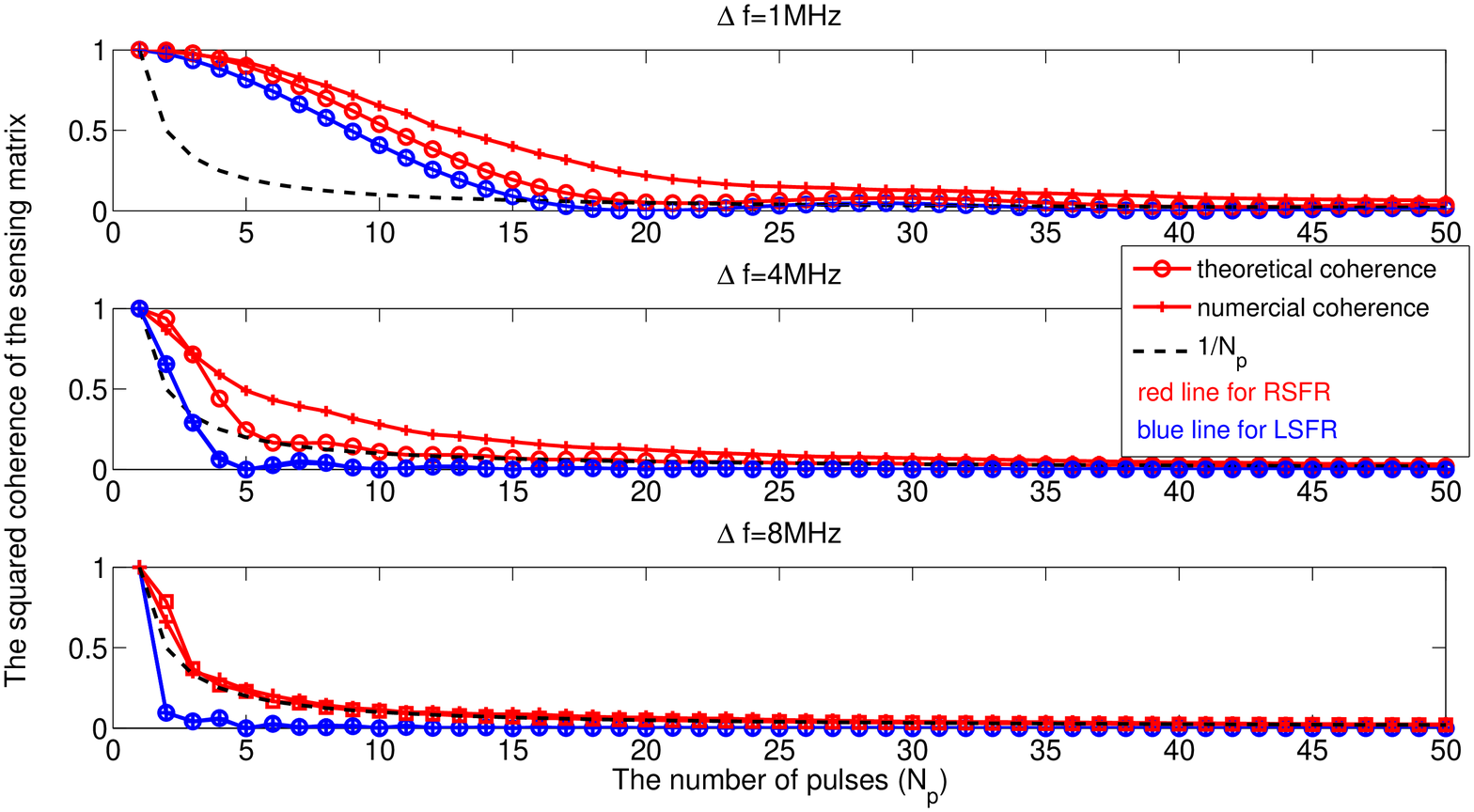}
\caption{  \footnotesize{Average squared coherence of the
sensing matrix
 for different numbers of pulses $N_p$ over $100$ independent and
random runs ($\Delta f=4\times 10^{6}$, $M=M_t=10$ and $N_r=1$). The
distance of two  grid points in the range plane is $\Delta
 c=7.5m$.
  }}\label{correlation_LSF_RSF}
\end{figure}

The goal of this subsection is to test the performance of CSSF MIMO
radar based on LSF and RSF.  Figure \ref{correlation_LSF_RSF}
compares the numerical and theoretical squared coherence
 of the sensing matrix corresponding to two adjacent grid points in the range plane for different numbers of
 pulses and various values of the linear frequency step $\Delta
 f=1M Hz$, $4M Hz$ and $8M Hz$.
  All the results shown in Fig. \ref{correlation_LSF_RSF} are
 the  numerical squared coherence averaged over 100 independent and random runs and the theoretical squared coherence
 for LSFR and RSFR calculated  based on (\ref{constant_step}) and
 (\ref{random_step}).
  We consider the  case in which  $M_t=M=10$, $N_r=1$ and the grid step is $\Delta
  c=7.5m$.
 For a fair comparison, we choose   random step frequencies within the  same frequency band  as in LSFR, i.e., $f+[0, (N_p-1)\Delta
 f]$.
It can be easily seen that the numerical  squared coherence of the
sensing matrix for LSFR  perfectly matches with the theoretical
results in (\ref{constant_step}). The numerical  squared coherence
of the sensing matrix for RSFR approaches the theoretical results in
(\ref{random_step}) as $\Delta f$ increases and  approaches $1/N_p$
as the number of pulses increases. It is also verified by Fig.
\ref{correlation_LSF_RSF} that  LSFR exhibits  lower coherence of
the sensing matrix than does RSFR.

\begin{figure}[htbp]
  \centering
    \includegraphics[height=3in,width=4in,clip=true]{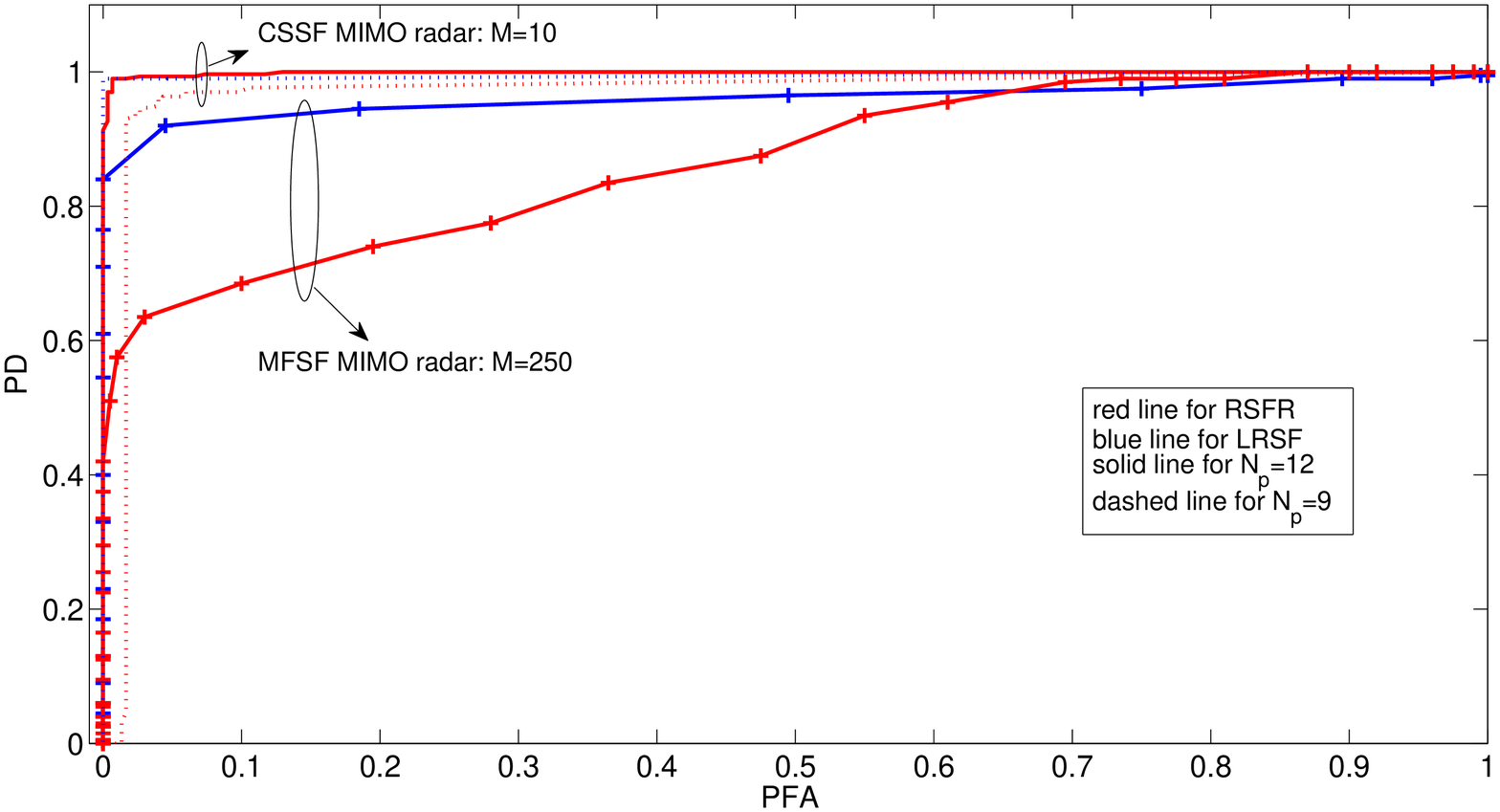}
    \caption{\footnotesize{ROC of range estimates obtained with  linearly and randomly stepped frequency  CSSF MIMO radar and MFSF MIMO radar ($M=M_t=10$, $N_r=1$ and $\Delta f=1MHz$).
  }}\label{range_estimation}
\end{figure}

  Figure \ref{range_estimation} shows
the receiver operating characteristic (ROC) curves of the range
estimates produced by the random and linear step-frequency technique
based on 200 random and independent runs.  Here, the probability of
detection (PD) is the
 percentage   of cases in which all the targets are detected. The probability of false alarm (PFA) is the percentage of
cases in which false targets are detected.
 We consider a case in which the angle and speed of three targets  are the same and assumed to be
 known. In each independent  run, the target angle and speed
 are randomly generated.
 The ranges of three targets are  fixed to $1005 m$, $1010 m$ and $1045 m$,
 respectively. The power of the
jammer signal is $4$ and SNR$=0$dB.
We can see that the use of  LSF yields better performance than
randomly choosing the carrier frequency within the same frequency
band.
 In this particular case, CS-based RSFR  requires 12 pulses to generate the ROC performance that can be achieved by   CS-based LSFR using only $9$
pulses.
 The performance of LSFR and RSFR based on the MFM is also shown in
 Fig. \ref{range_estimation}.
 It can be seen that the former using $12$ pulses   is far better than the latter with the same number of  pulses. It can also be seen that
  CSSF MIMO radar outperforms  MFSF MIMO radar.

\subsection{The joint angle-Doppler-range estimation of CSSF MIMO radar}
\begin{figure}[htbp]
  \centering
    \includegraphics[height=2.8in,width=3.5in,clip=true]{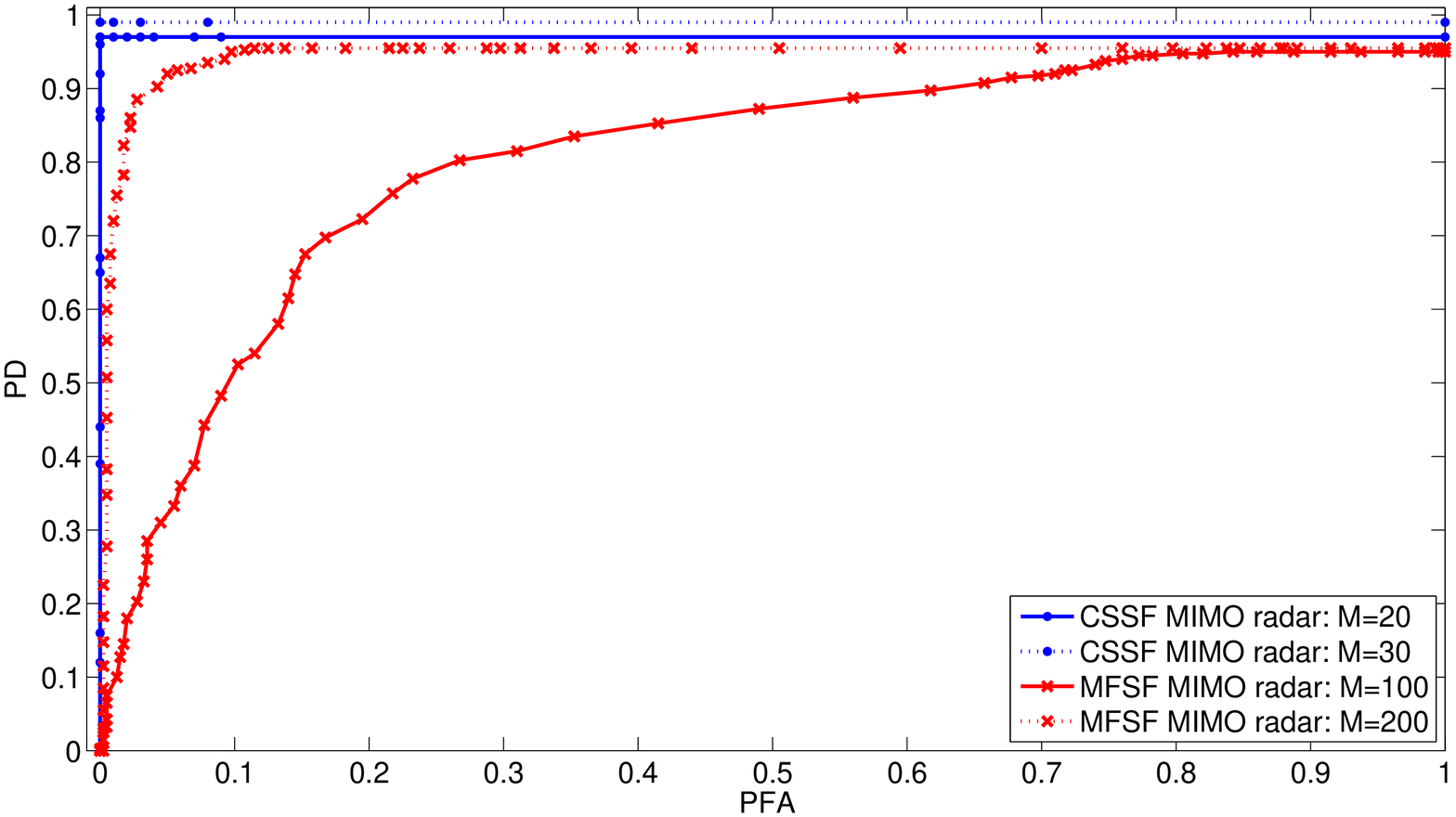}
 \caption{\footnotesize{  ROC of target detection based on angle-speed-range estimates yielded by
 the proposed  decoupled scheme in Section \ref{decoupled_scheme} for CSSF MIMO radar  and MFSF MIMO radar
 ($M_t=10,N_r=\tilde{N_r}=7$ and $N_p=12$).}
  }\label{ROC}
\end{figure}

  Figure \ref{ROC} shows
the ROC curves of the angle-speed-range estimates yielded by CSSF
MIMO radar using the
 decoupled scheme. The angle-speed-range
estimates have been obtained based on $200$ random and independent
runs. The cases in which  $M_t=10$, $N_r=\tilde{N_r}=7$ and $N_p=12$
are shown in Fig. \ref{ROC}.
 The azimuth angle and range of three  targets are
randomly generated in each run but the spacing of angle and range
between targets are   fixed to $0.3\textordmasculine $ and $7.5m$,
respectively. The speeds of three targets are  $10m/s, 30m/s,$ and $
60m/s$.  The power of the jammer signal is $4$ and SNR$=0$dB. The
performance of MFSF MIMO radar, shown in Fig. \ref{ROC}, is obtained
in the same  decoupled fashion, i.e., 1) estimate target angle and
range based on a single pulse; then refine the angle estimates based
on  the finer angle grid points around the initial angle estimates
by using the MFM; and then 2) process the entire pulse train to
extract angle-speed-range information by discretizing the speed
space, constructing  finer range grid points around the initial
range estimates and
  utilizing the initial
angle estimates obtained in 1).  One can see that MFSF MIMO radar is
inferior to CSSF MIMO radar even when using far more measurements
than the latter.

\section{Conclusions}
We have presented a  CSSF MIMO radar system that applies SF to
CS-based MIMO radar. The technique of SF can significantly improve
range resolution. We have shown that CSSF MIMO radar has the
potential to achieve better resolution than  MFSF MIMO radar, and
that more pulses are required by RSFR than by LSFR to achieve the
desired performance  with all other parameters being the same. The
angle-Doppler-range estimation requires discretization of the
angle-Doppler-range space into a large number of grid points, which
would increase the complexity of the CS approach.
 We have presented
a  CSSF MIMO radar scheme that  by decoupling angle-range estimation
and Doppler estimation achieves significant complexity reduction.
The proposed technique applies to   slowly moving targets and relies
on  initial rough angle-range estimates. Assuming that the initial
estimates do not miss any targets, the proposed low complexity
scheme maintains the high resolution of the CS approach.

\bigskip
\centerline{\bf Acknowledgment} The authors would like to thank Dr.
Rabinder Madan of the Office of Naval Research for sharing his ideas
on the use of compressive sampling in the context of MIMO radar.

\bibliographystyle{IEEE}

\appendices

\end{document}